\documentclass[aps,twocolumn,superscriptaddress]{revtex4}
\usepackage{epsfig}
\usepackage{times}
\begin{document}

\title{A double-edged sword: Benefits and pitfalls of heterogeneous punishment in evolutionary inspection games}

\author{Matja{\v z} Perc}
\email{matjaz.perc@uni-mb.si}
\affiliation{Faculty of Natural Sciences and Mathematics, University of Maribor, Koro{\v s}ka cesta 160, SI-2000 Maribor, Slovenia}
\affiliation{Department of Physics, Faculty of Sciences, King Abdulaziz University, Jeddah, Saudi Arabia}
\affiliation{CAMTP -- Center for Applied Mathematics and Theoretical Physics, University of Maribor, Krekova 2, SI-2000 Maribor, Slovenia}

\author{Attila Szolnoki}
\email{szolnoki@mfa.kfki.hu}
\affiliation{Institute of Technical Physics and Materials Science, Research Centre for Natural Sciences, Hungarian Academy of Sciences, P.O. Box 49, H-1525 Budapest, Hungary}

\begin{abstract}
As a simple model for criminal behavior, the traditional two-strategy inspection game yields counterintuitive results that fail to describe empirical data. The latter shows that crime is often recurrent, and that crime rates do not respond linearly to mitigation attempts. A more apt model entails ordinary people who neither commit nor sanction crime as the third strategy besides the criminals and punishers. Since ordinary people free-ride on the sanctioning efforts of punishers, they may introduce cyclic dominance that enables the coexistence of all three competing strategies. In this setup ordinary individuals become the biggest impediment to crime abatement. We therefore also consider heterogeneous punisher strategies, which seek to reduce their investment into fighting crime in order to attain a more competitive payoff. We show that this diversity of punishment leads to an explosion of complexity in the system, where the benefits and pitfalls of criminal behavior are revealed in the most unexpected ways. Due to the raise and fall of different alliances no less than six consecutive phase transitions occur in dependence on solely the temptation to succumb to criminal behavior, leading the population from ordinary people-dominated across punisher-dominated to crime-dominated phases, yet always failing to abolish crime completely.
\end{abstract}

\maketitle

In $1982$ Wilson and Kelling \cite{WILSON_AM82} introduced the ``broken windows theory'', explaining how seemingly unimportant and harmless signals of urban disorder may over time elicit antisocial behavior and serious crime. The central premise of the theory is simple yet powerful, and it is reminiscent of preferential attachment or the Matthew effect \cite{rigney_13, perc2014matthew} with a negative connotation. Just like the more connected nodes attract more new links during network growth \cite{barabasi_s99, ALBERT_RMP01}, so does an unattended broken window invite bypassers to behave mischievously or even disorderly. Similarly, a graffiti might point to an unkept environment, signaling that more egregious damage will likely be tolerated as well. One broken window is thus likely to become many broken windows, and the inception of urban decay and criminal behavior is in place.

The simplicity of this widely adopted criminological theory invites mathematicians and physicists to adopt a complex systems approach \cite{gell1988simplicity} to study criminal behavior \cite{orsogna15}, in particular since the collective behavior of the system in this case can hardly be inferred from the relatively simple individual actions. Emergent phenomena such as pattern formation including percolation \cite{wang_z_srep12, yang_hx_njp14} and phase transitions are commonly associated with complex social and biological systems \cite{szabo_pr07, CASTELLANO_RMP09, pacheco_c14, szolnoki_jrsif14}, and in this realm the mitigation of crime is certainly no exception. Recent research highlights that crime is far from being uniformly distributed across space and time \cite{ALVES_PONE13, PICOLI_SR14}, and this is confirmed also by the dynamic nucleation and dissipation of crime hotspots \cite{SHORT_MMMAS08, SHORT_PNAS10, RODRIGUEZ_MMMAS10, BERESTYCKI_MMS13} and the emergence of complex criminal networks \cite{HEGEMANN_PA11, CATANESE_SNAM13, FERRARA_ESA14, DUIJN_SR14}.

The emergence of crime can also be treated as a social dilemma \cite{santos_pnas06, NOWAK_S06, rand_tcs13}, in as far that social order is the common good that is threatened by criminal activity, with competition arising between criminals and those trying to prevent crime. An adversarial evolutionary game with four competing strategies has recently been proposed \cite{SHORT_PRE10}, where paladins are model citizens that do not commit crimes and collaborate with authorities, while villains, at the other extreme of the spectrum, commit crimes and do not report them. Intermediate figures are informants who report on other offenders while still committing crimes, and apathetics who neither commit crimes nor report to authorities. Apathetics are similar to second-order free-riders in the context of the public goods game with punishment \cite{FEHR_N04, SIGMUND_TREE07, PANCHANATHAN_N04, helbing_ploscb10}, in that they cooperate at first order by not committing crimes, but defect at second order by not punishing offenders. Simulations have revealed that in the realm of the adversarial game informants are key to the emergence of a crime-free society, and this has subsequently been confirmed also with human experiments \cite{DORSOGNA_PONE13}.

In general, the mitigation of crime can be framed as an evolutionary game with punishment, although recent research has raised doubts on the use of sanctions as a means to promote prosocial behavior \cite{DREBER_N08, HERRMANN_S08, RAND_S09, RAND_NC11, vukov_pcbi13}. Rewards for not doing and reporting crime are a viable alternative, and in this case the ``stick versus carrot'' dilemma becomes an important consideration \cite{HILBE_PRSB10, SZOLNOKI_NJP12, szolnoki_prx13, BERENJI_PONE14}. In the context of rehabilitating criminals, the question is also how much punishment for the crime and how much reward for eschewing wrongdoing in the future is in order for optimal results, as well as whether these efforts should be the responsibility of individuals or institutions \cite{gurerk_s06, sigmund_n10, szolnoki_pre11} under the assumption of a limited budget \cite{chen_xj_njp14}.

It is at this intersection of statistical physics of complex system and evolutionary games that we aim to contribute in the present paper by considering a three-strategy spatial inspection game with uniform punishment as well as a five-strategy spatial inspection game with heterogeneous punishment. The inspection game is a recognized model in the sociological literature for the dynamics of crime \cite{becker_jpe68, tsebelis_rs90}. The game addresses the question of why anybody would be willing to invest into costly punishment of criminals, given that individuals are tempted to benefit from the punishing activities of others without actively contributing to them. As soon as ordinary people are introduced who neither commit crimes nor contribute to their mitigation, one is thus faced with the second-order free-rider problem \cite{PANCHANATHAN_N04, perc_pone13}. As we will show in what follows, this may introduce cyclic dominance that enables the coexistence of all three competing strategies in the uniform punishment model. More importantly, the consideration of heterogeneous punisher strategies drastically elevates the complexity of possible solutions, revealing on the one hand a more effective solution to the second-order free-rider problem, yet still failing to abolish crime completely. As a consequence, the diversity of punishment allows the formation of different alliances between competing strategies, which gives rise to a sophisticated range of solutions in dependence on the payoffs.

In the next Section we first present the details of the considered 3-strategy and 5-strategy spatial inspection game, and then demonstrate how systematic Monte Carlo simulations reveal the benefits and pitfalls of punishing criminal behavior. Simulation details are described in the Methods Section. We conclude by discussing the presented results and their wider implications.

\section*{Results}

\subsection*{3-strategy and 5-strategy spatial inspection game}
We first introduce a three-strategy version of the spatial inspection game, where in addition to criminals $C$ and punishers $P$, also ordinary people $O$ compete for space on a $L \times L$ square lattice with periodic boundary conditions. We use the latter as the simplest network to account for the fact that the interaction range among individuals in human societies is limited. The payoff matrix\\

\begin{tabular}{r|c c c}
 & \,\,\,\,\,\,$O$ & \,\,\,\,\,\,\,\,$C$ & \,\,\,\,\,\,\,$P$\\
\hline
$O$ & \,\,\,\,\,\,0 & \,\,\,\,$-\beta$ & \,\,\,\,\,\,\,0\\
$C$ & \,\,\,\,\,\,$\beta$ & \,\,\,\,\,\,0 & \,\,\,\,\,\,$\beta-1$\\
$P$ & $-\alpha$ & \,\,\,\,\,\,$\gamma-\alpha$ & \,\,\,\,\,$-\alpha$\\
\end{tabular}
\vspace{3mm}

\noindent contains $\alpha$ as the punishment cost, $\beta$ as the temptation to succumb to criminal behavior as well as the loss when being a victim of crime, and $\gamma$ as the reward for punishing criminals. Moreover, when a criminal faces a punisher, it will receive $\beta-1$, where $-1$ corresponds to the normalized punishment fine. These payoffs apply for each pairwise interaction between the players.

To enable a more sophisticated response to the second-order free-rider problem, we also consider an extended model with heterogeneous punishment. Similarly to other diversity-motivated social problems \cite{perc_pre08, santos_n08, santos_jtb12}, we expect that such a model will provide further insights and a more adequate answer to the free-rider problem. In the proposed five-strategy version of the spatial inspection game punishers are divided into three categories, namely $L$, $M$ and $H$, depending on the cost they are willing to bear for punishing criminals. The extended payoff matrix\\

\begin{tabular}{r|c c c c c}
 & \,\,\,\,\,\,$O$ & \,\,\,\,\,\,\,\,$C$ & \,\,\,\,\,\,\,$L$ & \,\,\,\,\,\,\,$M$ & \,\,\,\,\,\,\,$H$\\
\hline
$O$ & \,\,\,\,\,\,\,\,0 & \,\,\,\,$-\beta$ & \,\,\,\,\,\,\,0 & \,\,\,\,\,\,\,0 & \,\,\,\,\,\,\,0\\
$C$ & \,\,\,\,\,\,\,\,$\beta$ & \,\,\,\,\,\,0 & \,\,\,\,\,\,$\beta-\frac{1}{3}$ & \,\,\,\,\,\,$\beta-\frac{2}{3}$ & \,\,\,\,\,\,$\beta-1$\\
$L$ & $-\frac{1}{3}\alpha$ & \,\,\,\,\,\,$\frac{1}{3}(\gamma-\alpha$) & \,\,\,\,\,$-\frac{1}{3}\alpha$ & \,\,\,\,\,$-\frac{1}{3}\alpha$ & \,\,\,\,\,$-\frac{1}{3}\alpha$\\
$M$ & $-\frac{2}{3}\alpha$ & \,\,\,\,\,\,$\frac{2}{3}(\gamma-\alpha$) & \,\,\,\,\,$-\frac{2}{3}\alpha$ & \,\,\,\,\,$-\frac{2}{3}\alpha$ & \,\,\,\,\,$-\frac{2}{3}\alpha$\\
$H$ & \,\,\,\,$-\alpha$ & \,\,\,\,\,\,\,\,\,\,\,$\gamma-\alpha$ & \,\,\,\,\,\,\,\,\,$-\alpha$ & \,\,\,\,\,\,\,\,\,$-\alpha$ & \,\,\,\,\,\,\,\,\,$-\alpha$\\
\end{tabular}
\vspace{3mm}

\noindent contains the same three main parameters as the three-strategy payoff matrix, with the key difference being that punishers $L$ and $M$ are willing to bear only $1/3$ and $2/3$ of the full punishment cost $\alpha$, respectively. Naturally, they also receive a proportionally smaller reward $\gamma$. Punishers $H$ correspond to punishers $P$ in the three-strategy model in terms of their commitment to sanctioning criminals, but we introduce a different notation for convenience.

Both the uniform three-strategy and the heterogeneous five-strategy spatial inspection game are studied by means of Monte Carlo simulations, as described in the Methods section.

\subsection*{Evolutionary dynamics}

\begin{figure}
\centerline{\epsfig{file=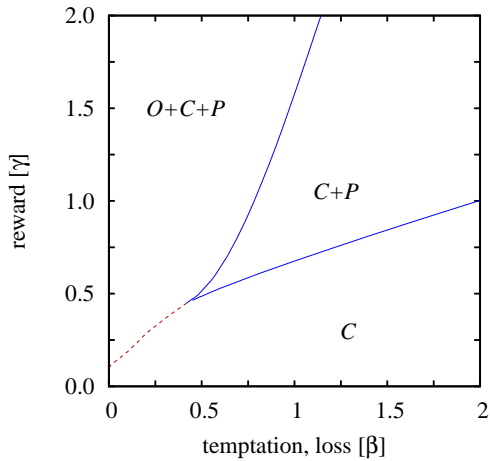,width=7cm}}
\caption{Phase diagram of the three-strategy spatial inspection game with uniform punishment. Depicted are strategies remaining on the square lattice after sufficiently long relaxation times as a function of the temptation/loss $\beta$ and the reward for punishing criminals $\gamma$, as obtained for the the punishment cost $\alpha=0.5$. Here $C$ marks the parameter region where the population terminates in a homogeneous ``all-criminal'' phase, $C+P$ marks the region where criminals and punishers coexist, while in the $O+C+P$ region all three strategies are present in the stationary state due to cyclic dominance. Solid blue lines denote continuous phase transitions, while the dashed red line denotes the border of cyclic dominance between competing strategies.}
\label{phase}
\end{figure}

We begin by presenting the complete $\beta-\gamma$ phase diagram at a representative value of the punishment cost $\alpha$ in Fig.~\ref{phase}. It can be observed that criminals dominate if the reward for their punishment $\gamma$ is small. If the reward exceeds a certain value at a fixed  temptation/loss $\beta$, then the punishers become viable. At moderate $\beta$ values, however, their presence is also accompanied by the emergence of ordinary players. The stability of the $O+C+P$ phase is due to cyclic dominance between the three competing strategies \cite{szolnoki_jrsif14}. In particular, within the $O+C+P$ region ordinary people outperform the punishers, the punishers defeat the criminals, while the criminals beat ordinary people, thus closing the $O \to P \to C \to O$ loop of dominance. Conversely, for larger values of $\beta$, in particular if $\beta>\alpha$, the pure $C$ phase becomes the two-strategy $C+P$ phase via a second-order continuous phase transition as $\gamma$ increases. Moreover, at sufficiently large values of the reward $\gamma$, the three-strategy $O+C+P$ phase and the two-strategy $C+P$ phase are separated by a second-order continuous phase transition.

\begin{figure}
\centerline{\epsfig{file=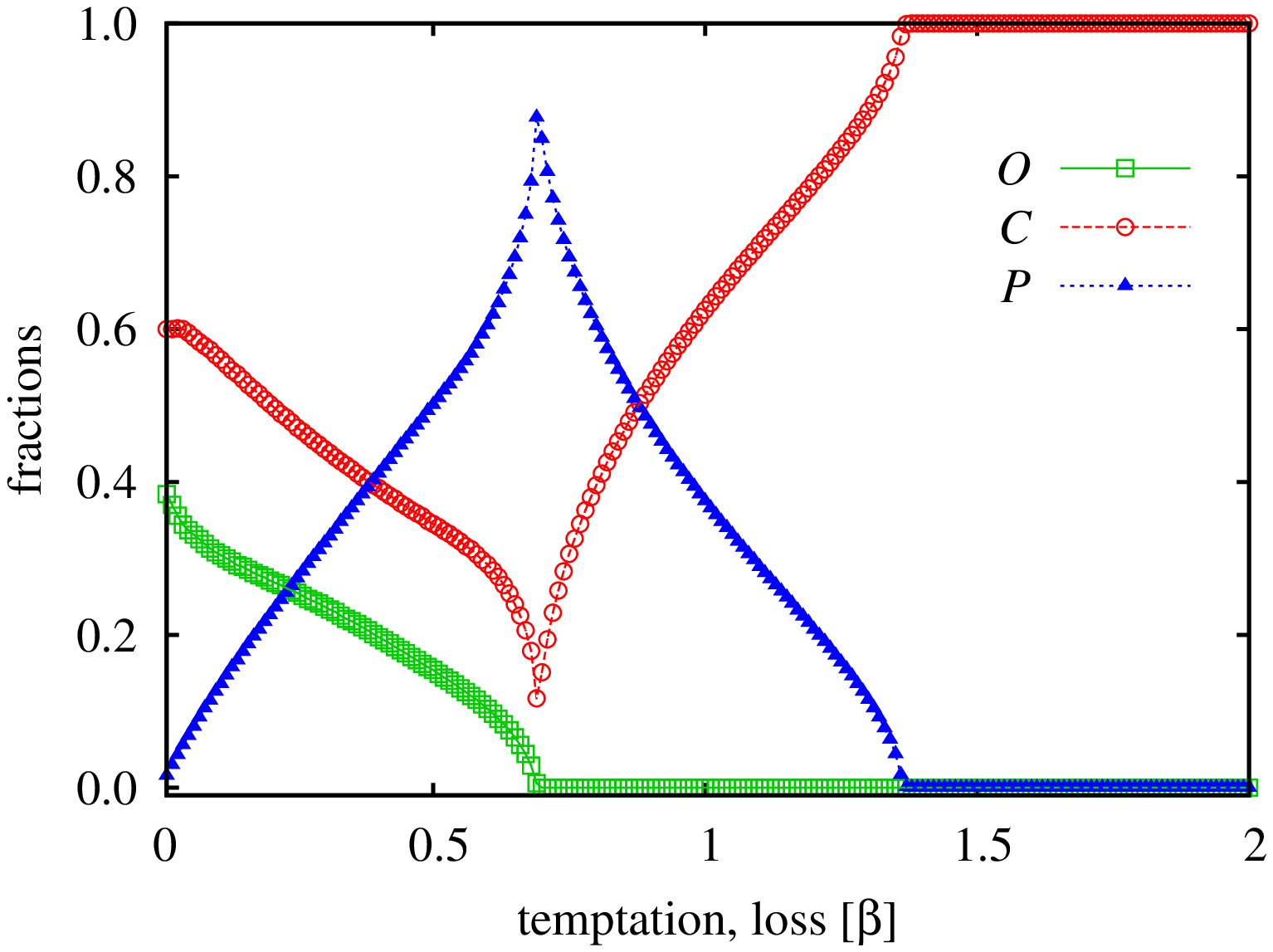,width=8.5cm}}
\centerline{\epsfig{file=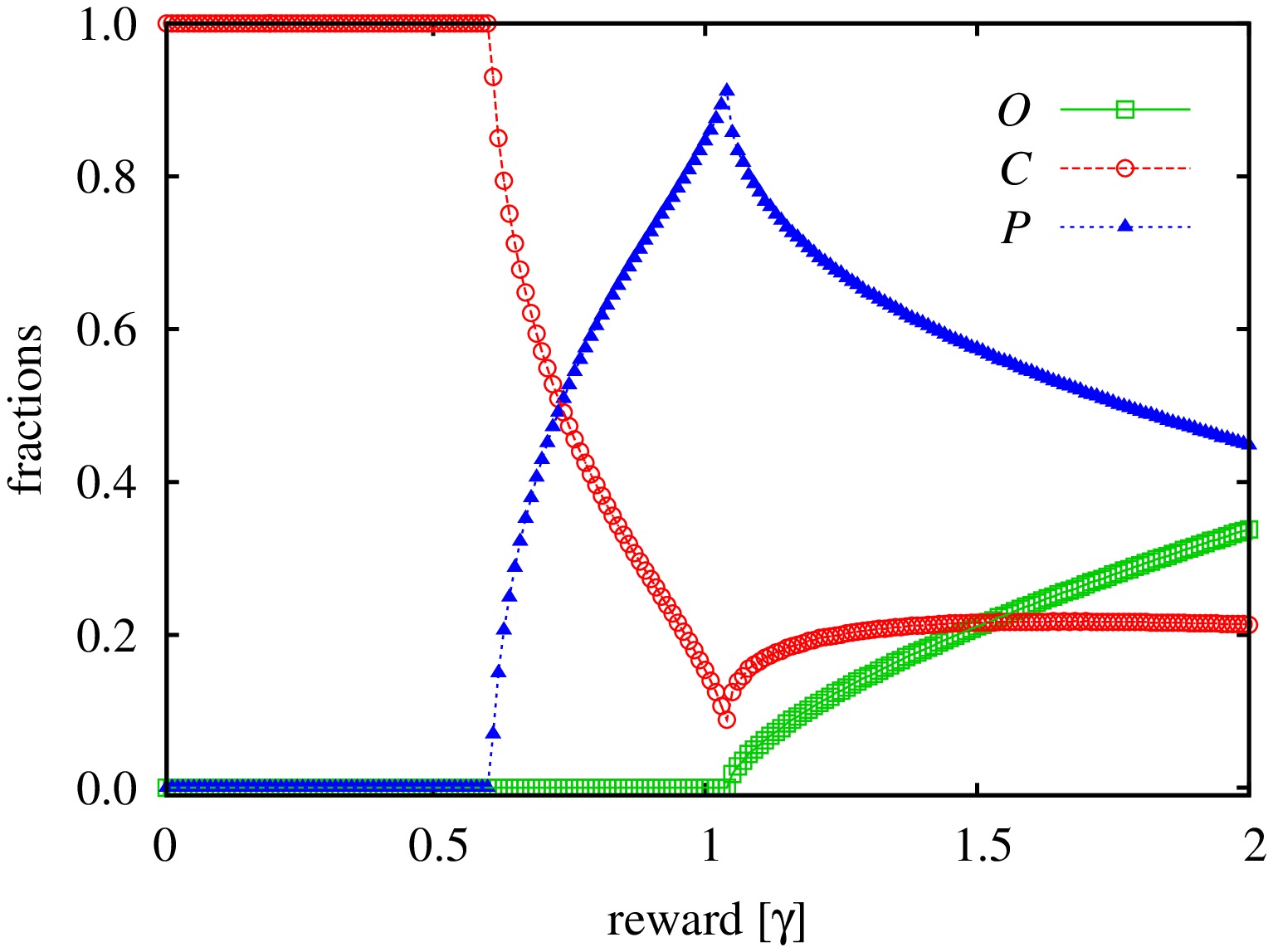,width=8.5cm}}
\caption{Two characteristic cross-sections of the phase diagram depicted in Fig.~\ref{phase}. Left panel shows the fraction of the three strategies in dependence on the temptation/loss $\beta$ at $\gamma=0.8$. Starting at the three-strategy $O+C+P$ phase, the fraction of ordinary people and the criminals decreases steadily with increasing the value of $\beta$ until eventually $O$ die out and the two-strategy $C+P$ phase is reached. Immediately thereafter the fraction of criminals starts rising as the value of $\beta$ increases further, with the second continuous phase transition marking the emergence of the pure $C$ phase. Right panel shows the fraction of the three strategies in dependence on the reward for punishing criminals $\gamma$ at $\beta=0.8$. In this case we start at the pure $C$ phase, which turns to the two-strategy $C+P$ phase as soon as $\gamma$ is large enough to sustain the punishers. As $\gamma$ increases further ordinary people become viable too through a second continuous phase transition, ultimately yielding the three-strategy $O+C+P$ phase that is maintained by cyclic dominance. In both panels the punishment cost is $\alpha=0.5$.}
\label{cross}
\end{figure}

For a more quantitative view, we present in Fig.~\ref{cross} characteristic cross-sections of the phase diagram shown in Fig.~\ref{phase}. These cross-sections confirm that criminals can dominate in the high temptation/loss region or in the low reward region. Moreover, it can be observed that larger rewards are beneficial for the punishers, but only up to a certain point. If $\gamma$ increases beyond a critical point ordinary people emerge, and as second-order free-riders they flourish on the expense of those that punish criminal behavior. We emphasize that, interestingly, the payoffs of ordinary people are independent of $\gamma$, yet still their fraction increases as $\gamma$ increases. This counterintuitive result is due to cyclic dominance, where feeding the prey, in this case the punishers who do get larger payoffs for larger $\gamma$ values, directly benefits the predator, which in this case are the ordinary people \cite{frean_prsb01, szabo_jpa04}. We can thus conclude that the real obstacle in the fight against criminal behavior is the possibility of ordinary people to free-ride on the efforts of punishers. A similar conclusion has been reached before for the evolution of cooperation in the public goods game with punishers, where the free-riding problem of defectors is simply deferred to the second-order free-riding problem of cooperators \cite{FEHR_N04}.

\begin{figure}
\centerline{\epsfig{file=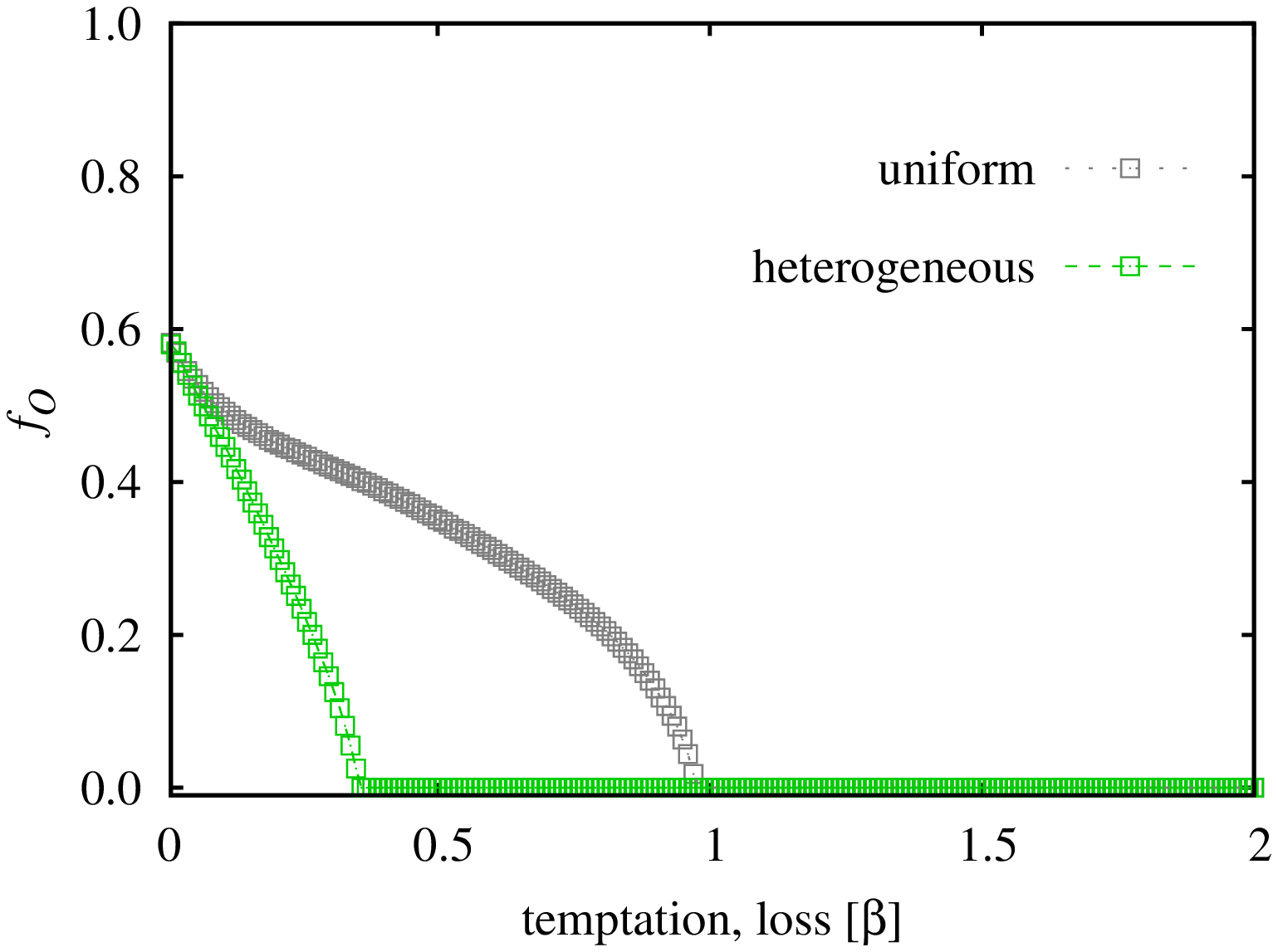,width=8.5cm}}
\centerline{\epsfig{file=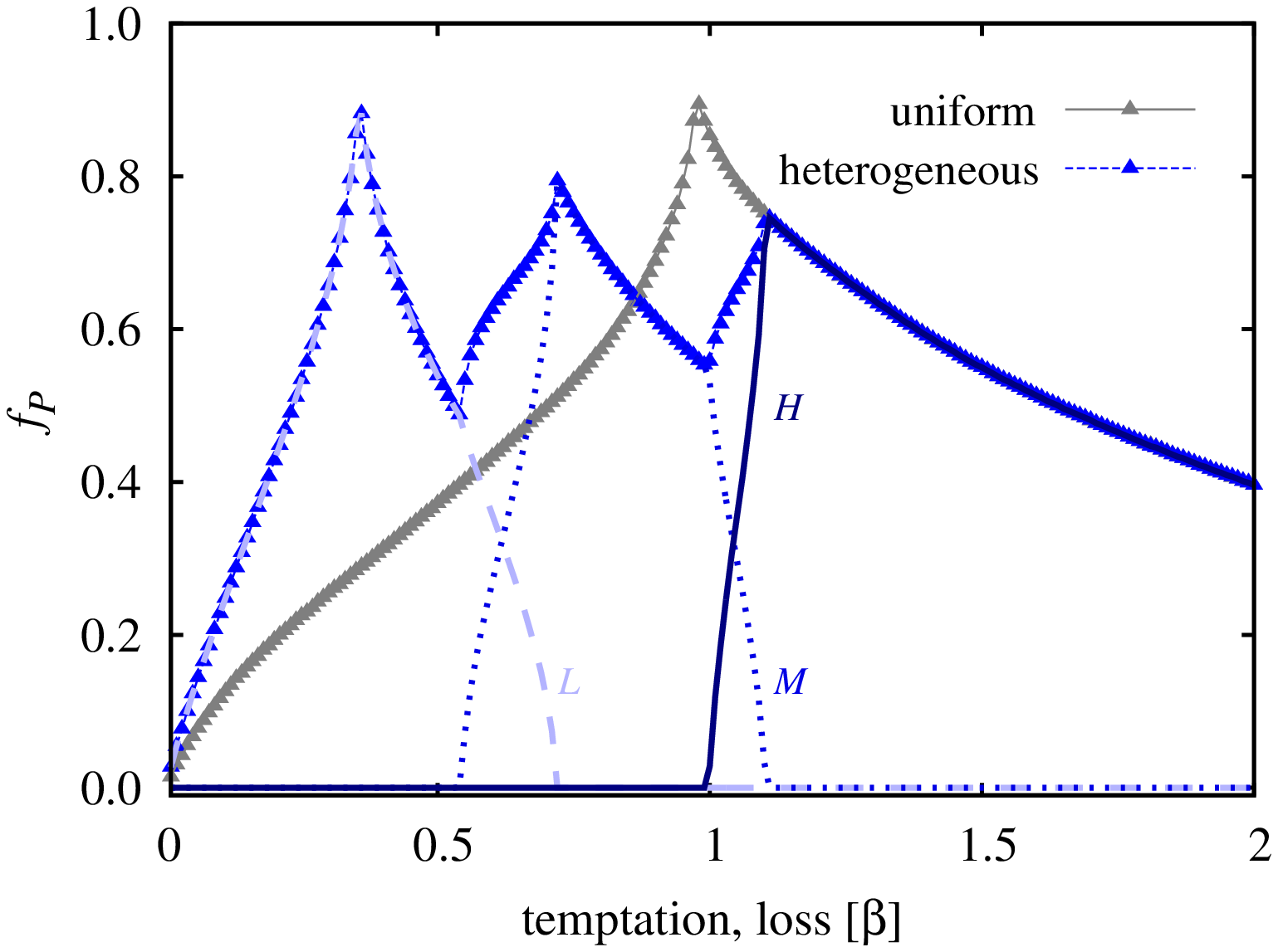,width=8.5cm}}
\caption{Left panel shows the fraction of ordinary people in dependence on the temptation/loss $\beta$, as obtained for the three-strategy spatial inspection game with uniform punishment and the five-strategy spatial inspection game with heterogeneous punishment (see legend). It can be observed that heterogeneous punishment is indeed more effective in eliminating second-order free-riding by ordinary people than uniform punishment. Right panel shows the fraction of punishers in dependence on the temptation/loss $\beta$ for the uniform punishment model and the aggregate fraction of all punishers in the heterogeneous punishment model, as well as the fraction of punishers $L$, $M$ and $H$ individually (see legend). The success of heterogeneous punishment to eliminate second-order free-riding is somewhat relativized, as higher punishment levels will not necessarily lead to lower criminal levels (see Fig.~\ref{cext} for an explanation). The origin of the zig-zag outlay of the aggregate fraction of all punishers is analyzed in Fig.~\ref{snapshots}.
In both panels the punishment cost is $\alpha=0.5$ and the reward for punishing criminals is $\gamma=1.5$.}
\label{opext}
\end{figure}

As a natural response of punishers to the harmful exploitation of ordinary people, we next consider the five-strategy spatial inspection game with heterogeneous punishment. In particular, strategies $L$ and $M$ try to eschew the exploitation by reducing the amount they contribute for sanctioning to $1/3$ and $2/3$ of the full cost, respectively. However, their reward is proportionally smaller as well (see the extended payoff matrix in Section 2 for details). Due to the large number of competing strategies and the resulting multitude of possible subsystem solutions we focus on the most important parameter region where ordinary players survive in the uniform, three-strategy, model. Accordingly, we explore a representative cross section when the reward is high enough for punishing strategies to survive, and we explore how the system responds to the diversity of punishment.

\begin{figure}
\centerline{\epsfig{file=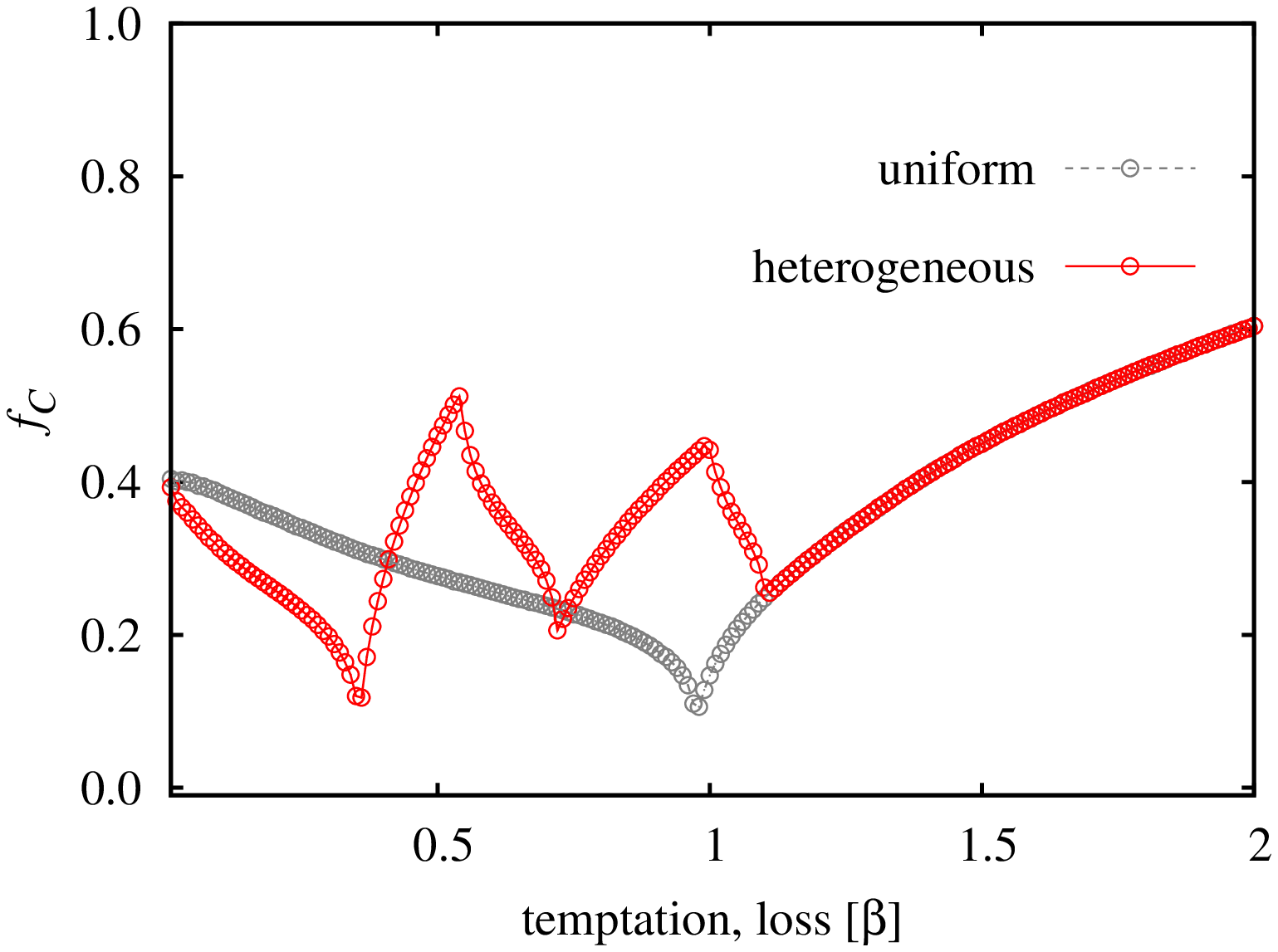,width=8.5cm}}
\centerline{\epsfig{file=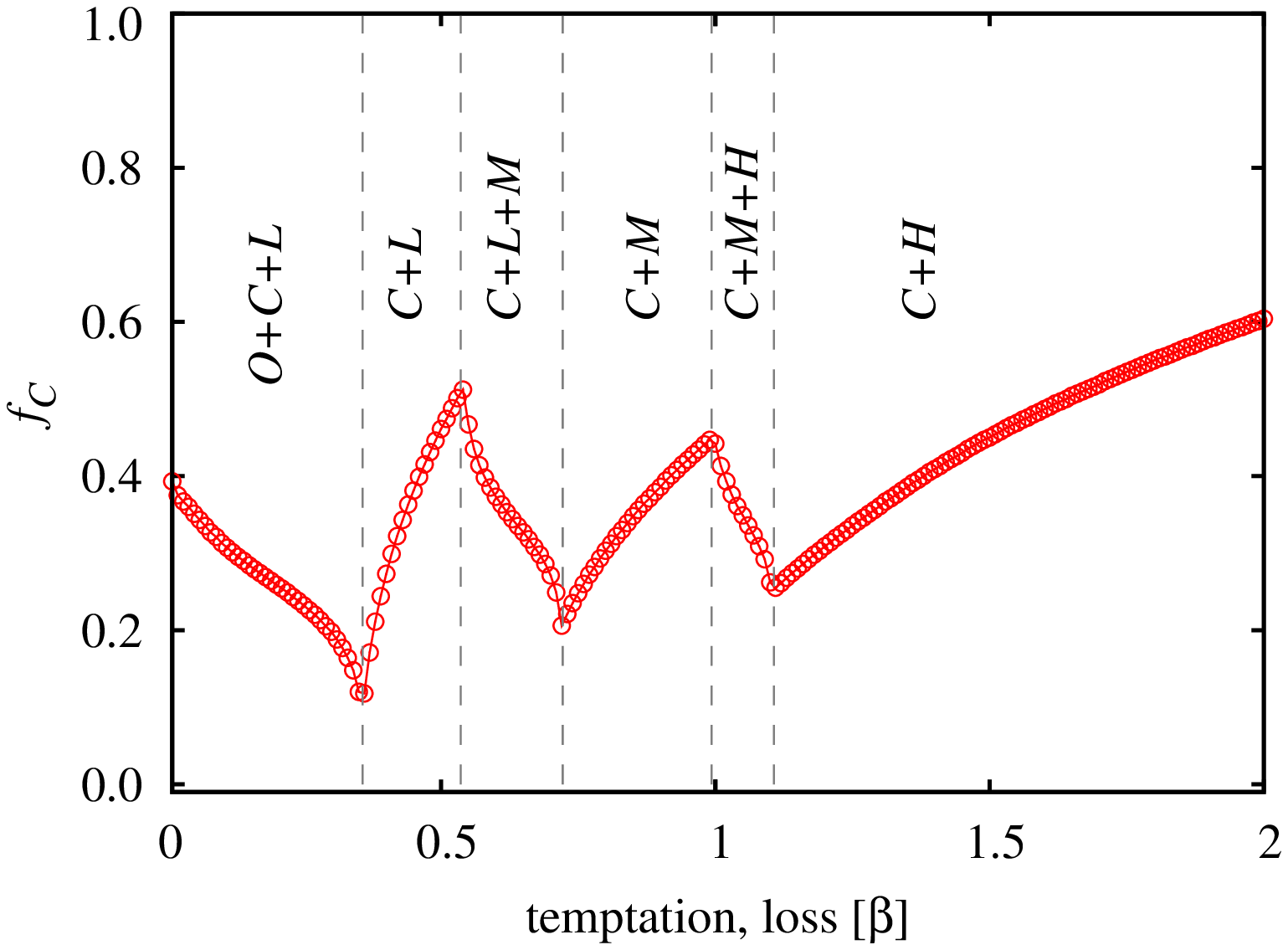,width=8.5cm}}
\caption{Top panel shows the fraction of criminals in dependence on the temptation/loss $\beta$, as obtained for the three-strategy spatial inspection game with uniform punishment and the five-strategy spatial inspection game with heterogeneous punishment (see legend). It can be observed that heterogeneous punishment is more effective than uniform punishment in eliminating crime only in the low $\beta$ limit, which also agrees with the region in which second-order free-riding is deterred more efficiently (see Fig.~\ref{opext}). In general, however, uniform punishment works just as well or better than heterogeneous punishment in abating crime. Bottom panel again shows the fraction of criminals, along with the different phases that contain the $C$ strategy. Despite the multitude of consecutive phase transitions in dependence on solely a single parameter, criminal behavior is never completely eliminated. In both panels the punishment cost is $\alpha=0.5$ and the reward for punishing criminals is $\gamma=1.5$.}
\label{cext}
\end{figure}

Results presented in the left panel of Fig.~\ref{opext} confirm the effectiveness of resorting to heterogeneous punishment in that second-order free-riders are able to survive only in a significantly narrower interval of the temptation/loss $\beta$ if compared to the uniform punishment model. Furthermore, results presented in the right panel of Fig.~\ref{opext} also give credence to the expectation that the reduced viability of ordinary people will promote the evolution of punishers. More precisely, we find that the uniform punishment strategy is significantly less effective than heterogeneous punishment for almost the entire range of the temptation/loos $\beta$, except for a narrow interval in the $\beta>\alpha$ region. As we will show in Fig.~\ref{cext}, this fact has important consequences for the mitigation of criminal behavior in the population.

\begin{figure*}
\centerline{\epsfig{file=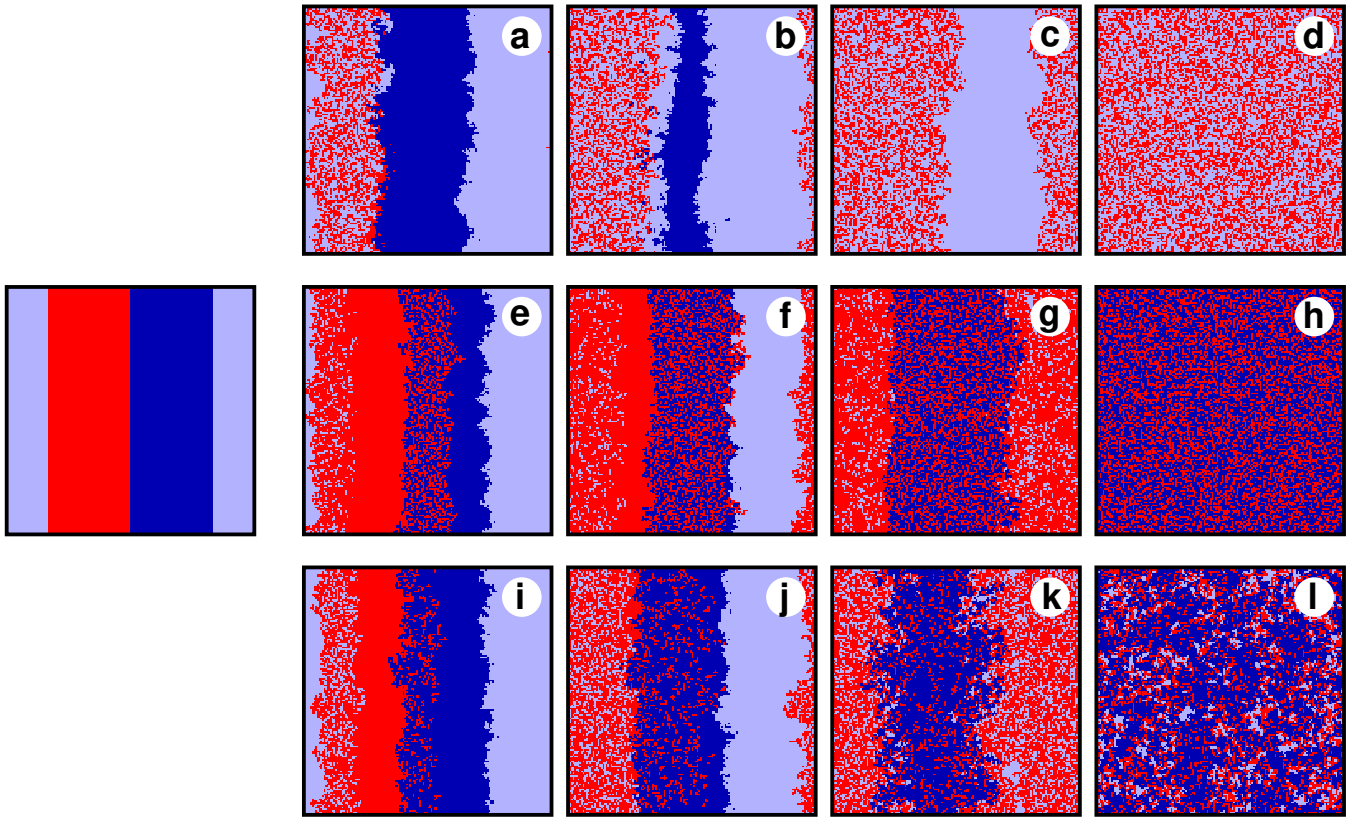,width=18cm}}
\caption{Time evolution of strategy distributions in the population, as obtained with the heterogeneous punishment game starting from the same prepared initial state (leftmost panel) for $\gamma=1.5$ and three different values of the temptation/loss: (a)-(d) $\beta=0.5$, (e)-(h) $\beta=0.9$, and (i)-(l) $\beta=0.7$. The resulting three different stationary states are reached within 400 MCS, which are depicted in the rightmost panels. Colors red, light blue and dark blue depict the location of $C$, $L$ and $M$ players, respectively. For visual clarity, we have used a small $150 \times 150$ system size. See main text for a detailed description of the different evolutionary outcomes.}
\label{snapshots}
\end{figure*}

Another peculiarity that can be observed in the right panel of Fig.~\ref{opext} is the zig-zag outlay of the aggregate fraction of all punishers in the five-strategy model. Yet this can be understood thoroughly simply by looking at the fraction of punishers $L$, $M$ and $H$ individually. The mentioned panel reveals clearly that low values of $\beta$ are able to sustain only those punishers who are willing to invest the lowest cost towards sanctioning criminals. The rank of the most viable punishers subsequently increases from $L$ over $M$ to $H$ as we increase $\beta$, and the solution of the five-strategy model thus eventually becomes identical to the the solution of the three-strategy model. Remarkably, we can observe six consecutive phase transitions [$(O+C+L) \to (C+L) \to (C+L+M) \to (C+M) \to (C+M+H) \to (C+H) \to C$] as we increase a single parameter, $\beta$. It is worth pointing out that the reported increment of the punisher rank with increasing the temptation/loss $\beta$ resonates with the outcome of a recent human experiment \cite{jiang_ll_pone13}, where, in the realm of a social dilemma, it was shown that if cooperation is likely one should punish mildly.

We continue with the results presented in Fig.~\ref{cext}, where we compare the effectiveness of uniform and heterogeneous punishment to deter criminal behavior. To a degree unexpected, it can be observed that the possibility to resort to different levels of punishment does not necessarily work better than uniform punishment in reducing crime. On the contrary, the fraction of $C$ players is generally higher over a large interval of $\beta$ values when the heterogeneous punishment model is used. More precisely, the fraction of criminals is lower only in the low temptation/loss region where $L$ punishers can adjust to this favorable condition. This observation is related to the failure of heterogeneous punishment to eliminate second-order free-riding more effectively than uniform punishment, and it indicates that sophisticatedly adjusted punishers may win a battle against ordinary people, but loose the main war against the actual enemy, the criminals. While punishers can lower the amount they invest towards sanctioning criminals, such a reduced effort also yields smaller rewards. Interestingly, the positive side of lower costs can be utilized only if the heterogeneity of punishers is maintained. The said effect becomes visible if we mark the borders of different phases on the curve of criminals, as shown in the right panel of Fig.~\ref{cext}. As it is illustrated, the fraction of criminals can be a decaying function even if we increase the temptation/loss $\beta$, but only as long as different types of punishers exist and compete against the criminals. As soon as evolution favors a single punisher type, an effective response to an increase of the value of $\beta$ becomes absent. Lastly, we note that the conclusions attained with the results presented in Figs.~\ref{opext} and \ref{cext} remain generally valid also for all high temptation values.

To obtain a better understanding of the origin of the zig-zag outlay of criminals depicted in Fig.~\ref{cext}, we monitor the time evolution of the distribution of strategies in the population for three different combinations of payoff parameters, as shown in Fig.~\ref{snapshots}. We emphasize that the main mechanism responsible for the formation of different stationary states is due to the different motion of interfaces that separate the possible solutions of the system. Accordingly, we follow the evolution of interfaces starting from a prepared initial state, but for clarity only two types of punishers are present because this minimal model is sufficient to capture the essence of the emerging effect. The extrapolation to the full five-strategy model, however, is straightforward. For comparison, we use an identical prepared initial state, as shown in the leftmost panel, for three representative values of $\beta$. As in previous figures, red color depicts $C$ players while light and dark blue depict the $L$ and $M$ punishers, respectively. Before discussing each specific case, we note that, individually, $L$ always beats $M$ due to the lower cost of inspection. When the temptation/loss is low, as shown in panels (a)-(d), $M$ can beat $C$ very efficiently, while $L$ is unable to do the same but simply coexists with the criminals. The superiority of $L$ over $M$, however, will result in a shrinking area of the $M$ domain, as shown in panel (b). Ultimately, this fact leads to the extinction of strategy $M$, despite the fact that it is more successful in deterring criminals than strategy $L$. As soon as $M$ die out, as shown in panel (c), criminals can exploit the milder punishment from strategy $L$ and spread towards the stationary state, as shown in panel (d). A seemingly surprising and counterintuitive result is that criminals, who can coexist with $L$ players but are defeated by $M$ players, are able to survive while their ``predators'' ($M$) go extinct. But in fact, the evolution depicted in the panels (a)-(d) simply illustrates the actual consequence of second-order free-riding. Namely, $L$ players exploit the more altruistic $M$ players by contributing less to sanctioning criminals. In the absence of $L$ players, however, the common enemy ($C$) can spread relatively free and reach a significantly high level ($f_C \approx 0.46$).

Interestingly, when $M$ players are less successful in deterring $C$ players, the outcome is completely the opposite, as shown in panels (e)-(h) of Fig.~\ref{snapshots}. Since the temptation/loss $\beta=0.9$, $C$ are able to coexist with $M$. The coexistence of $C$ and $L$ strategies is also still possible, and at the same time $L$ continue to invade the pure $M$ phase [the invasion ends in panel (f)]. However, $L$ become ineffective against the $C+M$ alliance. Indeed, this two-strategy alliance is so powerful that it beats the other $C+L$ alliance completely. The competition between the two alliances starts in panel (g), and it terminates with the total victory of the $C+M$ alliance in panel (h). The conclusion is similar as in the preceding case. Namely, when the evolution selects only one type of punishers, then criminals have a reasonable chance to survive. Note that the fraction of criminals in the stationary state is again relatively high, $f_C \approx 0.40$, despite of substantial punishment.

The most favorable outcome can be obtained at an intermediate temptation/loss value, as shown in panels (i)-(l) of Fig.~\ref{snapshots}. The $\beta=0.7$ value is still high enough to maintain the coexistence of the $C+M$ alliance, but it lessens its evolutionary advantage in that the $C+L$ alliance is able to survive. The stationary state thus contains three strategies, whereby a relatively small portion of the population, $f_C \approx 0.27$, is occupied by criminals. We thus conclude that, in the long-run, if different punisher strategies survive in the stationary state, heterogeneous punishment may be utilized successfully to mitigate crime better than uniform punishment. Note that $f_C$ is a decreasing function of $\beta$ in the three-strategy phase in Fig.~\ref{cext}, while it always increasing when homogeneous punishment is applied (in $C+L$, $C+M$, or in the $C+H$ phases). This is because heterogeneous punishment enables the validation of the most effective approach against crime: sometimes moderate efforts, yielding milder fines, serve the interest of whole population better than severe punishment. Even more importantly, the simultaneous presence of different types of punishers enables a synergy among them in that one strategy (in our case $M$) can lower the payoff of criminals significantly while the other strategy ($L$) can still enjoy a more competitive payoff due to a smaller cost. This multi-point effect is conceptually similar to when the duty of punishment is shared stochastically among cooperative players \cite{chen_xj_njp14}. Of course, as we have already emphasized, these conclusions remain valid and can be extrapolated to a larger number of different punisher strategies.

\section*{Discussion}
We have studied the effectiveness of punishment in abating criminal behavior in the spatial inspection game with three and five competing strategies, entailing criminals, ordinary people and punishers. In the five-strategy game, we have introduced three different types of punishers, depending on the amount they are willing to contribute towards sanctioning criminals. We have shown that cyclic dominance plays an important role in that it maintains the survivability of seemingly subordinate strategies through indirect support. For example, increasing the reward for punishing criminals might promote second-order free-riding of ordinary people, despite of the fact that it should in fact support the punishers. This is due to cyclic dominance, where directly promoting the prey, in this case the punishers, benefits the predator, which in this case are the ordinary people. Moreover, we have shown that the actual obstacle in the fight against criminal behavior is the possibility of ordinary people to free-ride on the efforts of punishers, which is also the main culprit behind the establishment of cyclic dominance. In general, sanctioning criminal behavior is thus a double-edged sword. The obvious benefit is that the evolution of crime is contained and is unable to dominate in the population. The pitfall is that, in conjunction with ordinary people, punishment creates conditions that support cyclic dominance, which prevents the complete abolishment of crime even if the sanctions are severe and effective.

In addition to these observations, we have shown that the possibility of heterogeneous punishment yields a highly ambiguous measure against criminal behavior. At specific parameter values it can happen that milder punishers play the role of second-order free riders, which ultimately prevents to eliminate crime completely [see panels (a)-(d) in Fig.~\ref{snapshots}]. Evidently, the reverse process is also possible in structured populations where the more altruistic punishers can separate from second-order free riders and win the indirect territorial battle \cite{helbing_ploscb10, helbing_njp10}. But in the realm of the studied inspection game, we have also observed that the diversity of punishers can yield a more favorable social outcome even as the temptation to do crime is growing. In the latter case, the simultaneous presence of different punishers provides an advantageous coexistence: some punishers ensure a higher fine to criminal players while other punishers can benefit from a lower cost due to a less intensive engagement. Importantly, neither of these two options is effective on its own right, but together they improve the effectiveness of combating crime.

Notably, the emergence of cyclic dominance due to strategic complexity has been reported before, for example in public goods games with volunteering \cite{szabo_prl02}, peer punishment \cite{hauert_s07, helbing_ploscb10, amor_pre11, bednarik_prsb14}, pool punishment \cite{sigmund_n10, szolnoki_pre11} and reward \cite{szolnoki_epl10, SZOLNOKI_NJP12}, but also in pairwise social dilemmas with coevolution \cite{szolnoki_epl09, szolnoki_pre10b}. Other counterintuitive phenomena that are due to cyclic dominance \cite{ni_x_pre10, wang_wx_pre11} include the survival of the weakest \cite{frean_prsb01, berr_prl09}, the emergence of labyrinthine clustering \cite{juul_pre13}, and the segregation along interfaces that have internal structure \cite{avelino_pre14}, to name but a few examples. Cyclical interactions are thus in many ways the culmination of evolutionary complexity \cite{szolnoki_jrsif14}, and we here show that they likely play a prominent role in deterring crime as well. However, while the beneficial role of cyclic dominance for maintaining biodiversity is undeniable, one has to concur that it is a rather unsatisfactory outcome in terms of fighting criminal behavior. That is the sort of diversity in behavior that human societies could happily do without, yet it seems that this is precisely the trap the current system has fallen into. Indeed, data from the Federal Bureau of Investigation (see Fig.~2 in Ref.~\cite{orsogna15}) indicate that crime, regardless of type and severity, is remarkably recurrent. Although positive and negative trends may be inferred, crime events between 1960 and 2010 fluctuate across time and space, and there is no evidence to support that crime rates are permanently decreasing. The search for more effective crime mitigation strategies is thus in order, in particularly for such where the permanent elimination of crime is not an a priori impossibility.

\section*{Methods}
For both the 3-strategy and the 5-strategy spatial inspection game the Monte Carlo simulation procedure is the same. Initially all competing strategies are distributed uniformly at random on the square lattice. We note, however, that the reported final stationary states are largely independent of the initial fractions of strategies. Subsequently, in agreement with the random sequential update protocol, a randomly selected player $x$ acquires its payoff $\Pi_x$ by playing the game pairwise with all its four neighbors. Next, player $x$ randomly chooses one neighbor $y$, who then also acquires its payoff $\Pi_y$ in the same way as previously player $x$. Once both players acquire their payoffs, player $x$ adopts the strategy $s_y$ from player $y$ with a probability determined by the Fermi function
\begin{equation}
W(s_y \to s_x)=\frac{1}{1+\exp[(\Pi_x-\Pi_y)/K]},
\end{equation}
where $K=0.5$ quantifies the uncertainty related to the strategy adoption process \cite{blume_l_geb93, szabo_pr07}. In agreement with previous works, the selected value ensures that strategies of better-performing players are readily adopted by their neighbors, although adopting the strategy of a player that performs worse is also possible \cite{vukov_pre06, szolnoki_pre09c}. This accounts for imperfect information and errors in the evaluation of the opponent.

Each full Monte Carlo step (MCS) consists of $L^2$ elementary steps as described above, which are repeated consecutively, thus giving a chance to every player to change its strategy once on average. We typically use lattices with $600 \times 600$ players, although close to the phase transition points up to $9000 \times 9000$ players had to be used in this case to avoid accidental extinctions, and thus to arrive at results that are valid in the large-size limit. The fractions of competing strategies $f$ are determined in the stationary state after a sufficiently long relaxation time lasting up to $10^5$ MCS. In general, the stationary state is reached when the average of the strategy fractions becomes time-independent. Moreover, to account for the differences in initial conditions and to further improve accuracy, the final results are averaged over up to $100$ independent runs for each set of parameter values.

\begin{acknowledgments}
This research was supported by the Slovenian Research Agency (Grant P5-0027), the Hungarian National Research Fund (Grant K-101490), and by the Deanship of Scientific Research, King Abdulaziz University (Grant 76-130-35-HiCi).
\end{acknowledgments}

\end{document}